\documentclass[12pt]{iopart}
\usepackage{graphicx}
\begin{document}

\title[Amplification and discrimination of photons]{A scheme for amplification and discrimination of photons}

\author{A R Usha Devi$^{1,2}$, R Prabhu$^{3,1}$ and A K Rajagopal$^{4,2}$}
\address{$^1$\ Department of Physics, Bangalore University, Bangalore-560 056, India}
\address{$^2$\ Inspire Institute Inc., McLean, VA 22101, USA}
\address{$^3$\ Department of Physics, Kuvempu University, Shankaraghatta, Shimoga-577 451, India}
\address{$^4$\ Center for Quantum Studies, George Mason University, Fairfax, VA 22030, USA}

\ead{arutth@rediffmail.com}

\begin{abstract}
A scheme for exploring photon number amplification and discrimination 
is presented  based on the interaction of a large number of two-level atoms with a single mode radiation field. 
The fact that the total number of photons and  atoms in the excited states is a constant under time evolution in 
Dicke model is exploited to rearrange the atom-photon numbers. Three significant predictions emerge from our 
study: Threshold time for initial exposure to photons, time of perception (time of maximum detection 
probability),  and discrimination of first few photon states.
\end{abstract}

\pacs{42.50.Ct, 03.65.Ta.}
\submitto{\JPB}
\maketitle

\section{Introduction}
A scheme for investigating amplification and discrimination of input photons is presented based on the 
interaction of a large number of identical two-level atoms and low intensity single mode radiation.  
This analysis provides a deeper understanding of the cumulative response of millions of atoms to few input 
photons within the basic model~\cite{Dicke} of atom-field interaction.   
 
Historically, collective behavior of $N$ two level atoms with single mode radiation, investigated by 
Dicke~\cite{Dicke} in 1954, has led to vast array of interesting physical phenomena in quantum optics and 
recently, in artificial 
condensed matter systems~\cite{Brandes}.  The present investigation is an entirely novel application of the 
Dicke model. The fact that the sum of the  number of photons and the number of atoms in the excited states is a 
constant under time evolution has a natural consequence of rearranging the atom-photon numbers 
resulting in photon  amplification as well as discrimination of small number of photons. A projective 
measurement on temporally evolving combined multiatom-radiation system  is shown to  lead to both these 
features.   

\section{Physical Model} 
The Hamiltonian characterizing the interaction of  $N$ two level atoms with a single mode  radiation is given by 
($\hbar=1$)
\begin{equation}
\label{dh}
H=\omega(a^\dag a)+\omega_0\, S_z+\frac{g}{\sqrt{N}}\,(S_+\, a+S_-\, a^\dag), 
\end{equation}
where $g$ denotes the atom-photon coupling parameter; $\omega_0$, the atomic splitting and $\omega$ the filed 
frequency;    $a^\dag\ (a)$ are creation (annihilation) operators of the field satisfying the bosonic 
commutation relations $[a,\, a^\dag]=1$. The collective (pseudo) spin operators of the two-state atoms 
\begin{equation}
S_{\pm}=S_x\pm i S_y=\frac{1}{2}\,\sum_{\alpha=1}^N\, \sigma_{\alpha \pm},\ 
S_z=\frac{1}{2}\,\sum_{\alpha=1}^N\sigma_{\alpha z}
\end{equation} 
 obey the commutation relations 
 \begin{equation}
 [S_+,S_-]=2S_z, \ \ [S_z,S_\pm]=\pm S_\pm. 
 \end{equation}          
It is well-known~\cite{TCM} that the excitation operator 
\begin{equation}
\label{constant}
N_{\rm ex}=a^\dag a + S_z+\frac{N}{2}
\end{equation} 
remains constant as the system evolves under the Hamiltonian (\ref{dh}). 
 The degenerate atom-photon  states  
$\vert S=\frac{N}{2},\, M=n_e-\frac{N}{2}\rangle \otimes \vert  n\rangle$ 
($-\frac{N}{2}~\leq~M~\leq~\frac{N}{2},$   or equivalently $n_e~=~ 0,1,2,\ldots , N,\ {\rm and}\  
n~=~0,1,2,\ldots $ ), with a fixed eigenvalue~\footnote{The  Fock states  $\vert n\rangle,\ n=0,1,2,\ldots$ 
denote eigenstates of the photon number operator $a^\dag a$ with eigenvalue $n$ and the Dicke states  
$\vert S=\frac{N}{2}, M\rangle,\ -\frac{N}{2}\leq M\leq\frac{N}{2}$ are joint eigenstates of 
$S^2=S_x^2+S_y^2+S_z^2$ and  $S_z$ with  eigenvalues $\frac{N}{2}(\frac{N}{2}+1)$,\  $M$ respectively.} 
$n+n_e$ of (\ref{constant}) span the space of the Hamiltonian.  
The symmetric atomic Dicke states $\vert S=\frac{N}{2},\, M=n_e-\frac{N}{2}\rangle$ (except for $n_e=0$ and $N$) 
are  well-known for their entanglement properties~\cite{Thiel,Bas,U}. These are essential in the ensuing 
discussion~\footnote{Dicke states with a specific permutation symmetry are chosen 
here  because of  the greatest simplicity offered by them in their theoretical 
analysis and also due to their current  experimental relevance~\cite{Thiel,Bas}.}.  The collective ground state 
of the 
atomic system is denoted by $\vert 0\rangle =\vert\frac{N}{2},\, -\frac{N}{2}\rangle$ throughout this sequel. 

 A given initial state of the combined atom-radiation system, $\rho_{AR}(0)$ evolves to 
 \begin{equation}
 \rho_{AR}(t)=e^{-iHt}\rho_{AR}(0)e^{iHt}.
 \end{equation}   
 Consequently, a projection operator $\Pi_{A0}\otimes I_R$, with $\Pi_{A0}=\vert 0\rangle\langle 0\vert,$  and 
$I_R$  the unit operator in the radiation space, gives us the conditional density matrix  of the collective 
atom-photon 
system ((where all the atoms are projected to their collective ground state)), 
subjected to the constraint (\ref{constant}). This projective measurement 
has the  implication that maximum number of photons are emitted consequently. The collective atomic ground state 
serves as a detection device and the projection operation at subsequent suitable intervals of time corresponds 
to efficient measurement of the radiation state containing maximum number of photons allowed.   It may be 
observed that  the constant of  motion (\ref{constant}) leads to rearrangement of the number of atoms in the 
excited state with the number of photons,  leading to photon amplification.  This is evident if $n_e>n$ number 
of atoms are initially in  excited state with $n$ input photons,  in which case $\langle N_{\rm ex}\rangle= 
n+n_e$ and  a subesquent  measurement projects the atoms to their collective ground state leading to 
$\langle a^\dag a\rangle(t) = n+n_e$.  More generally, $\langle a^\dag a\rangle(t)
=\langle N_{\rm ex}\rangle -\langle S_z\rangle(t) -\frac{N}{2}.$ A judicious choice of the initial state and the 
time of projection measurement is the basis of  photon amplification in this model. When a pure collective low 
lying excited state of atoms are considered, the method outlined above results in  photon number discrimination 
as well.              

\section{Photon amplification with different initial states} 
\subsection{Pure atom-photon state $\vert n_e;n\rangle$}   
Let us consider an initial atom-photon state  
\begin{equation}
\label{nen}
\vert S=\frac{N}{2}, M=n_e-\frac{N}{2}\rangle\otimes \vert n\rangle \equiv  \vert n_e; n\rangle, 
 \end{equation}
 with $n_e<<N,$ and the number of atoms $N$ sufficiently large so that the Holstein-Primakoff mapping~\cite{HP}
 \begin{equation}  
S_+=b^\dag\, \sqrt{N-b^\dag b}, \ S_-=\sqrt{N-b^\dag b}\, b,\ S_z=b^\dag b - \frac{N}{2}
\end{equation}  
in terms of bosonic operators $b,\, b^\dag$ satisfying 
\begin{equation}
[b,b^\dag]=1,
\end{equation} 
reduces  the Hamiltonian (\ref{dh}) into a two mode bosonic interaction structure~\cite{Brandes}  
(up to  order 
${\cal O}(1/N)$):
\begin{equation}
 \label{hinfty}
 H\sim\omega\, N_{\rm ex}-\omega_0\, \frac{N}{2}+g\, (b^\dag a+b\, a^\dag). 
 \end{equation} 

Temporal evolution in this approximation thus assumes the form, 
\begin{equation}
U(\tau)=e^{-iHt}=e^{-i\,\tau\, (\frac{\omega}{g}\, N_{\rm ex}-\frac{\omega_0}{2g}\, N)}
e^{-i\, \tau(b^\dag a+b\, a^\dag)}.
\end{equation}  
(Here we have denoted $gt=\tau$.) It may be noted that $U(\tau)$ acts as a passive unitary 
beam-splitter on the two mode bosonic Fock states $\vert n_e;n\rangle.$ Thus, initial states of the form 
(\ref{nen}) get confined  within the space spanned by the $(n_e+n+1)$ basis states $\vert n_e';n'\rangle$, 
with $n'_e,n'=0,1,2\ldots $ such that $n'_e+n'=n_e+n$ under time evolution.  
It is interesting to note that at scaled time $\tau=\pi/2$ the unitary operator $U(\tau)$  
swaps~\cite{xwang} the atom-photon numbers  
i.e., $\vert n_e; n\rangle\rightarrow\vert n; n_e\rangle$ with unit probability.         

In general, we have, 
\begin{eqnarray}
\label{tevol}
U(\tau)\vert n_e; n\rangle&=&e^{-i\tau[\frac{(n+n_e)\omega}{g} +\frac{N\omega_0}{2g}]}\, \sum_{n'_e,n'}\vert 
n_e',n'\rangle  
\langle  n_e',n'\vert e^{-i\tau (b^\dag a+b\, a^\dag)}\vert n_e,n\rangle \,  \nonumber \\ 
&=& e^{-i\tau\,[\frac{(n+n_e)\omega}{g} +\frac{N\omega_0}{2g}]}\,\, \sum_{n'_e,n'} \,   \vert n_e',n'\rangle\,  
d^{\frac{n_e+n}{2}}_{\frac{n'_e-n'}{2}, \frac{n_e-n}{2}}(2\tau)
 \nonumber \\
&& \hskip 0.5in \times \  e^{-i\pi[(n_e'-n')-(n_e-n)]/4}\,  \delta_{n_e'+n', n_e+n},   
\end{eqnarray}
where~\cite{Sak} 
\begin{eqnarray}
d^j_{m',m}(2\tau)&=&
\sum_k\frac{\sqrt{(j+m)!(j-m)!(j+m')!(j-m')!}}{(j+m+k)!k!(j-k-m')!(k-m+m')!}\, (-1)^{k-m+m'}\nonumber \\ 
&& \hskip 0.2in \times 
(\cos\tau)^{2j-2k+m-m'}\,(\sin\tau)^{2k-m+m'}=d^j_{m,m'}(-2\tau)
\end{eqnarray}
with the sum over $k$ taken such that none of the arguments of the factorials in the denominator are negative. 

A  measurement  $\Pi_{A0}\otimes I_R$ at an instant $t$  projects all the atoms to ground state: 
\begin{eqnarray}
\label{projstate}
&&\Pi_{A0}\otimes I_R\, U(\tau)\, \vert n_e; n\rangle =  
\sum_{n'}\, \vert 0,n' \rangle\ \   
d^{\frac{n_e+n}{2}}_{-\frac{(n+n_e)}{2},\frac{n_e-n}{2}}(2\tau)\, \delta_{n', n_e+n}\nonumber \\ 
&& \hskip 1.5in \times \ e^{i\pi[(n_e-n+n')]/4}\, 
e^{-i\tau\,[\frac{(n+n_e)\omega}{g}+\frac{N\omega_0}{2g}]}\nonumber \\
&&\hskip 0.8in=e^{-it(n\omega+\frac{N}{2}\omega_0)}\, 
e^{-in_e(\omega t-\frac{\pi}{2})}\,(-1)^{n_e}\, \sqrt{{\cal P}(n_e,n,\tau)}\ \, \vert 0;n+n_e\rangle.  
\end{eqnarray} 
  Thus, the probability of finding the atoms in ground state, which in turn corresponds to that of photon 
amplification $n\rightarrow n+n_e$, is given by 
\begin{equation}
\label{prob}
{\cal P}(n_e,n,\tau)=\left(\begin{array}{c} n+n_e \\ n_e\end{array}\right)\,
 \cos^{2n} (\tau)\, \sin ^{2n_e}(\tau).
 \end{equation}
\begin{figure}[h]
  \centering
\includegraphics[width=3in,keepaspectratio]{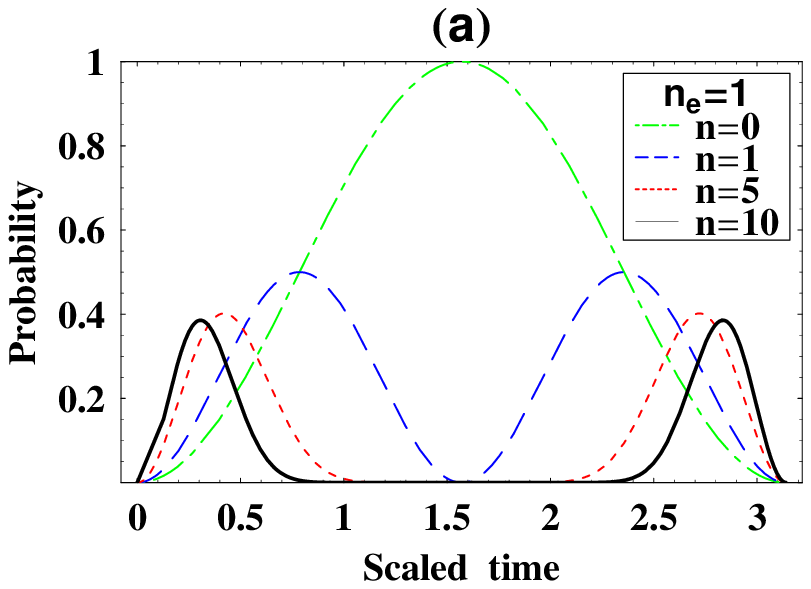}
\includegraphics[width=3in,keepaspectratio]{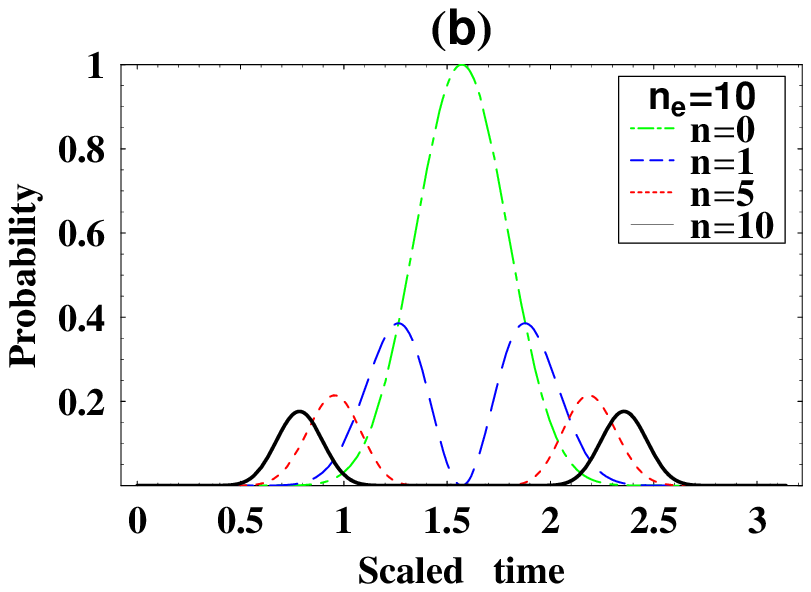}
\includegraphics[width=3in,keepaspectratio]{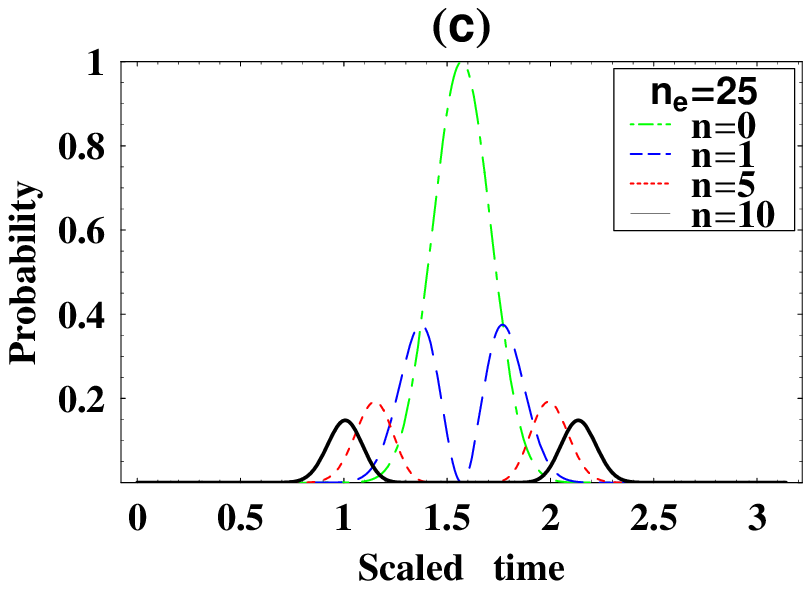}
  \caption{\label{fig1} Probability ${\cal P}(n_e,n,\tau)$ of finding the atoms in ground state as a function of 
scaled time $\tau$ for different values of initial photon numbers $n$.                 
   Maxima in probabilities occur at scaled 
times $\tau_n=\arccos\sqrt\frac{n}{n+n_e}.$ }
\end{figure}
In Figs.~\ref{fig1} (a), (b), (c), we have displayed the probability of maximum photon emission for three 
choices of  
$n_e=1,\ 10, $ and $25$, each with different values for the number of photons $n=0,1,5, 10,$ as a function of 
scaled time $\tau.$  For $n=0$ i.e., 
dark photon input state, the probability  has a single peak~\footnote{The probability ${\cal P}(n_e,n,\tau)$    
exhibits  repetitive structure with a period $\pi$, as is evident from (\ref{prob}).} in the range  
$0\leq \tau \leq \pi$ 
with maximum value ${\cal P}(n_e,0,\tau=\pi/2)=1$, which corresponds to  
swapping of atom-photon numbers $\vert n_e; 0\rangle\rightarrow\vert 0; n_e\rangle.$ 
However, when $n\neq 0$ the projective measurement $\Pi_{A0}\otimes I_R$ leads to 
$\vert n_e; n\rangle\rightarrow\vert 0; n+n_e\rangle$ corresponding to maximum photon emission; this is not a 
complete swap action and occurs with probability $0\leq {\cal P}(n_e,n\neq 0,\tau)<1$. Clearly, the probability 
${\cal P}(n_e,n\neq 0,\tau)$  vanishes at scaled time $\tau=\pi/2$, in order to give way to complete swap action  
$\vert n_e; n\rangle\rightarrow\vert n_e; n\rangle$. The probability profile for maximum photon emission 
$n\rightarrow n+n_e$ therefore reveals  two peaks  around $\tau=\pi/2$ for $n\neq 0$. 
The maxima of probabilities corresponding to different input photon numbers are well separated and they appear 
periodically at $\tau_n=\arccos\sqrt\frac{n}{n+n_e},$ allowing for photon number discrimination. 

We define the time of perception $\tau_p(n_e, n),$ for a given $n_e$ and $n$,  as the time at which the 
probability ${\cal P}(n_e,n,\tau)$ attains its  maximum. This determines the efficient detection of the photons. 
An examination of these figures reveal: (i) for a given $n_e$, the time of perception $\tau_p(n,n_e)$ reduces as 
more and more photons are detected - with lesser and lesser efficiency. Moreover the widths in the probabilities 
${\cal P}(n_e, n, \tau)$ reduce correspondingly. (ii) As $n_e$ increases, there is a threshold  time for the 
detection of photons, before which the probabilities are zero.   In particular, for $n_e=1$, there is an instant 
response to photons (for all $n$) as seen in Fig.~\ref{fig1}(a), whereas for higher values of $n_e$ there is a 
delay in such a response (See Fig.~\ref{fig1}. (b), and (c)).   (iii) For a given photon number $n$ the profile 
of probability as a function of time sharpens as $n_e$ increases; however maximum value of the probability drops 
with this, as is evident from Figs.~\ref{fig1}.

\subsection{Pure state of atoms with low intensity coherent radiation} 
The above discussion was confined to  pure 
photon number states. We now show that an enhanced photon amplification behavior is realized, when initially a 
low intensity  coherent  state of radiation 
\begin{equation}
\vert\alpha\rangle= e^{-\vert\alpha\vert^2/2}\,\sum_{n=0}^{\infty}\, \frac{\alpha^n}{\sqrt{n!}}\vert n\rangle,\ 
\vert\alpha\vert^2<1
\end{equation} 
is considered as input.  The probability of finding the atoms in ground state is obtained 
by following the procedures given in (\ref{tevol}) and  (\ref{projstate}):  
\begin{equation}
\label{probcoh}
{\cal P}(n_e,\alpha,\tau)=e^{-\vert\alpha\vert^2}\sum_{n=0}^{\infty}\,{\cal P}(n_e,n,\tau) 
\frac{\vert\alpha\vert^{2n}}{n!}.
\end{equation}
\begin{figure}[h]
\centering
 \includegraphics*[width=3in,keepaspectratio]{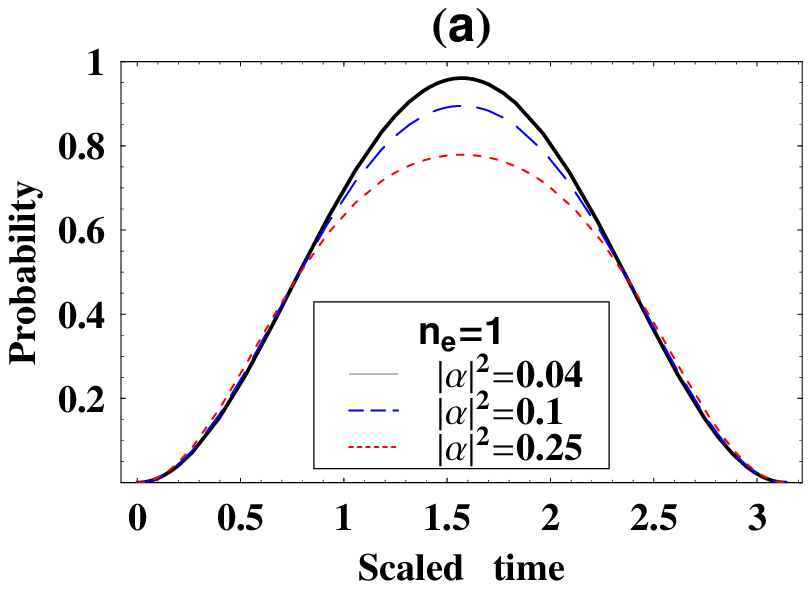}
 \includegraphics*[width=3in,keepaspectratio]{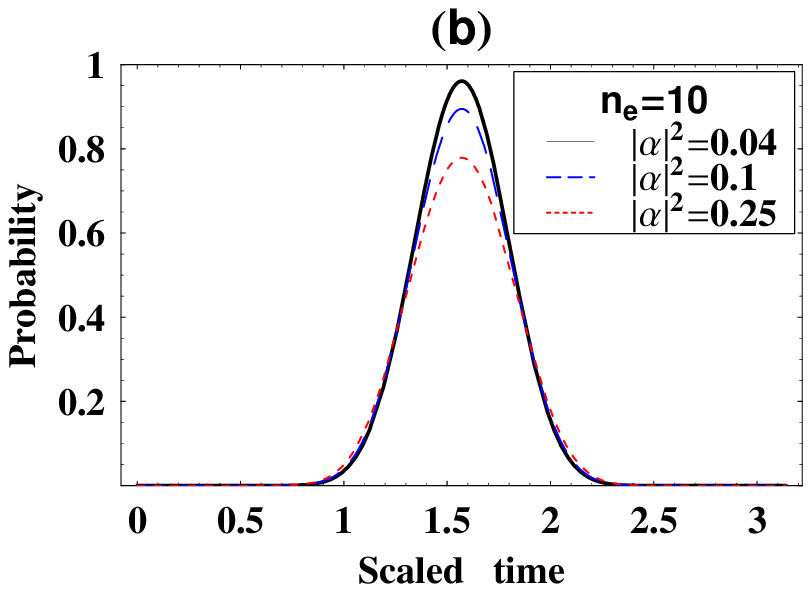}
 \includegraphics*[width=3in,keepaspectratio]{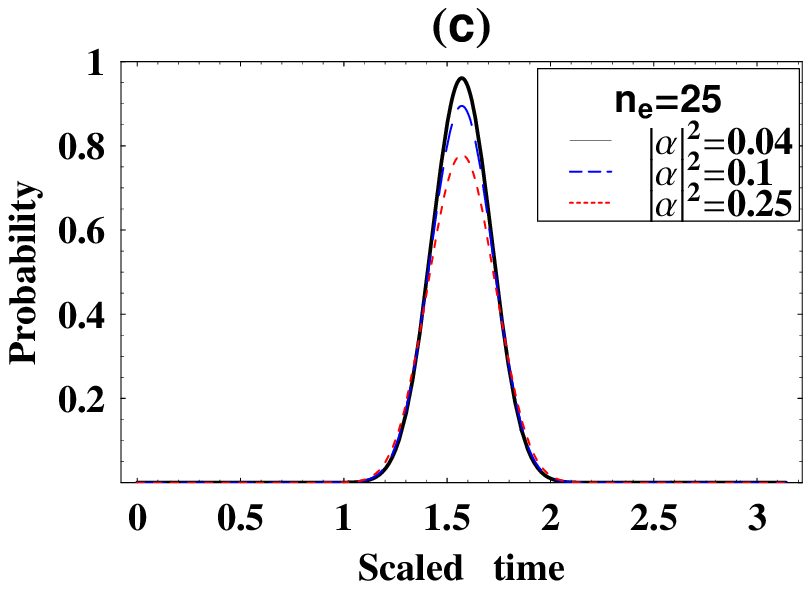}
 \caption{\label{fig2} Probability ${\cal P}(n_e,\alpha,\tau)$ of finding the atoms in ground state as a 
function of scaled time $\tau$ for different values of initial intensity $\vert\alpha\vert^2$ and for different 
choices of $n_e$, the initial number of atoms in excited state.}
\end{figure}

The probability profile (\ref{probcoh}) - with coherent radiation as input - is a series involving individual 
atom-photon Fock state probabilities ${\cal P}(n_e,n,\tau)$,   dark photon probability 
${\cal P}(n_e,0,\tau)$ being the leading term for low value of intensity $\vert\alpha\vert^2$. Thus, the 
probability response to low intensity coherent light has a similar structure as that of dark photon probability  
${\cal P}(n_e,0,\tau)$ (see Fig.~\ref{fig1}) with small contribution from higher order terms 
$\vert\alpha\vert^{2n}, \ n=1,2,\ldots$. It may be seen from Fig.~\ref{fig2} that the probability 
${\cal P}(n_e,\alpha,\tau)$ is found  to be nearly zero in the begining  and  raises to a maximum  at 
$\tau=\pi/2$  (peak value being 
${\cal P}(n_e,\alpha,\pi/2)\sim e^{-\vert\alpha\vert^2}$).    
An increase in the initial intensity of coherent radiation has the effect of reducing 
the probability of finding the atoms in their ground state. 
Amplification of the intensity of radiation, $\frac{I(t)}{I(t=0)}=\frac{\langle a^\dag 
a\rangle(t)}{\vert\alpha\vert^2},$  (after performing projective measurement $\Pi_{A0}$ on the temporally 
evolving state $U(\tau)\vert n_e; \alpha\rangle$)   
    approaches the value $\frac{n_e}{\vert\alpha\vert^2}$ as $\tau\rightarrow \frac{\pi}{2}$ i.e., when the 
probability of finding the atoms in ground state is maximum.     

\begin{figure}
\centering
 \includegraphics*[width=3in,keepaspectratio]{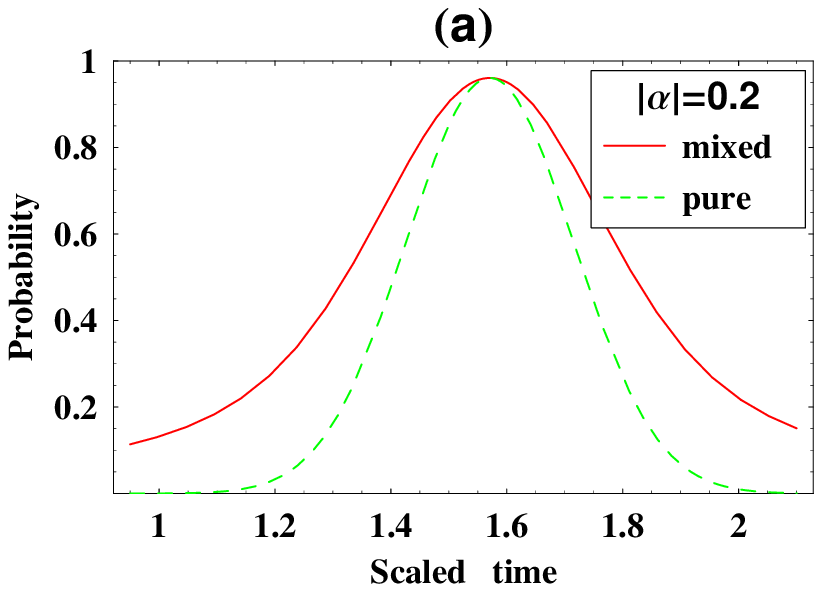}
 \includegraphics*[width=3in,keepaspectratio]{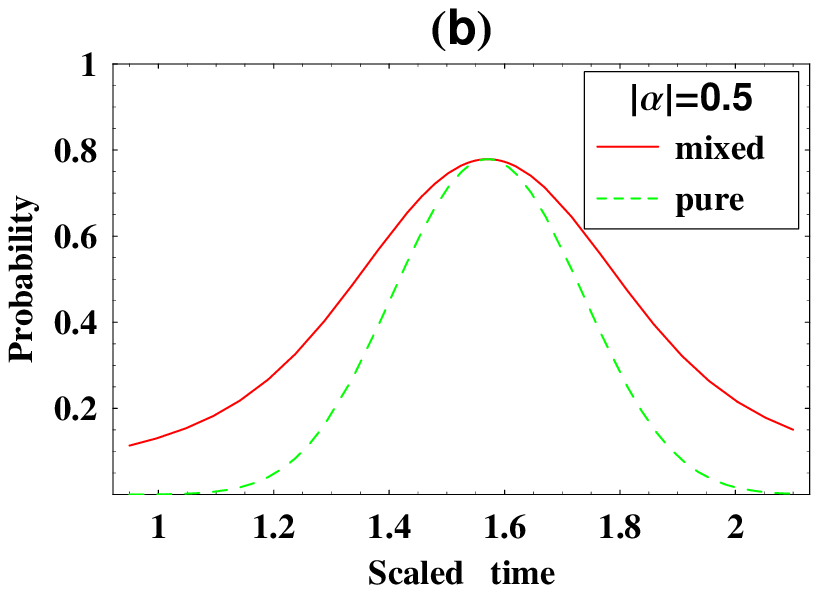}
 \caption{\label{fig3} Probability  of finding the atoms in ground state under time evolution governed by the 
Hamiltonian (\ref{hinfty}), when initial atomic state is chosen as (i) a mixed  state 
$\frac{1}{n_e+1}\,\sum_{m=0}^{n_e}\vert m\rangle\langle m\vert$, (ii) pure state $\vert n_e\rangle$, and the 
radiation in a coherent state, $\vert \alpha\rangle$,  as a function of scaled time $\tau$, for two different 
values of inital intensity of radiation and $n_e=25$. }
\end{figure}

\subsection{Mixed state of atoms with low intensity coherent radiation}
So far, we have considered pure atomic states. It is of interest to investigate how the effects found above may 
get modified when mixed  atomic states are employed. For simplicity, we choose here a chaotic mixture 
$(n_e+1)$ of  low lying collective excited states of atoms,  
\begin{equation}
\rho_{\rm atom}=\frac{1}{n_e+1}\,\sum_{m=0}^{n_e}\vert m\rangle\langle m\vert, \ \ n_e<<N,
\end{equation}
along with  low intensity coherent radiation. The probability of finding the atoms in ground state under time 
evolution in this model is readily found  to be, 
\begin{equation}
{\cal P}(\rho_{\rm 
atom},\alpha,\tau)=\frac{e^{-\vert\alpha\vert^2}}{n_e+1}\,\sum_{m=0}^{n_e}\sum_{n=0}^{\infty}\,\frac{\vert
\alpha\vert^{2n}}{n!}{\cal P}(m,n,\tau).
\end{equation}
The probability ${\cal P}(\rho_{\rm atom},\alpha,\tau)$ as well as ${\cal P}(n_e,\alpha,\tau)$ (see 
Eq.~(\ref{probcoh})) associated with a pure atomic state $\vert n_e\rangle$ are plotted, as a function of scaled 
time $\tau$,  in Fig.~\ref{fig3} (a) and (b) for two different choices of the initial intensity of radiation for 
a fixed 
$n_e=25.$ We observe that  while the maximum probability remains the same for a given initial intensity of 
radiation $\vert\alpha\vert^2,$  the mixed states $\rho_{\rm atom}$ result in a wider spread around the maximum 
value (at $\tau=\pi/2$)  compared to their pure state counterparts. Also, the probability ${\cal P}(\rho_{\rm 
atom},\alpha,\tau)$ builds above the background value $\frac{1}{n_e+1}.$   Moreover, as the intensity of the 
radiation  $\vert\alpha\vert^2$ increases, the maximum probability drops down. It may be noted that the photon 
amplification factor $\frac{I(t)}{I(0)}$ approaches the value $\frac{n_e}{2\, \vert \,\alpha\vert^2}$ as  
$\tau\rightarrow \pi/2$ for mixed atomic  states $\rho_{\rm atom}$ in contrast to its corresponding  value 
$\frac{n_e}{\vert\alpha\vert^2}$ in the case of  pure states $\vert n_e\rangle.$    

\section{Summary} We have presented a scheme for photon amplification and discrimination based on the Dicke 
model interaction between single mode radiation and $N$-atom system. This involves  Holstein-Primakopf large 
$N$  approximation~\cite{HP}, where the number $n_e$ of atoms in excited states is assumed to be much smaller 
than the total number $N$  of atoms. Further, this approximation enforces that the number $n$ of input photons 
is small, compared to $N$.   This  leads to a beam splitter feature 
for  time evolution  under the interaction Hamiltonian~(\ref{hinfty}), thus ensuring  entanglement between the 
atoms in collective excited states and the photons~\cite{beamsplitter}. Emission of maximum number of photons 
corresponds to a completely uncorrelated atom-radiation system, with all the atoms in the ground state and this 
is acheived via a projective measurement at a suitable interval of time. Concepts like  {\em threshold time}   
and  {\em time of perception} for exposure to low intensity light emerge as characteristic features of our 
investigation. Discrimination of  small number of photons is also realized in this scheme. This provides  
motivation for a quantum mechanical analysis of the collective response of millions of rods in the eye, which 
act as nearly perfect photon detectors, initiating the associated process of vision during 
night~\cite{Kan,RAKR,Reike}.

\section*{Acknowledgement} 
We thank the Referees for their insightful comments, which improved the presentation 
of our work in this revised version.

\end{document}